\documentclass[3p,11p]{elsarticle}

\usepackage{xcolor}
\usepackage{graphicx}
\usepackage{algorithm}
\usepackage{algpseudocode}
\usepackage{subfig}
\usepackage{float}
\usepackage{amssymb}
\usepackage{amsmath}
\usepackage{amsthm}
\usepackage{paralist}
\usepackage{ulem}
\usepackage{bbold}
\usepackage{bm}
\usepackage{comment}
\usepackage{booktabs}
\usepackage{xcolor}
\usepackage{soul}

\numberwithin{equation}{section}

\usepackage{mathtools}

\usepackage[title]{appendix}

\newtheorem{thm}{Theorem}

\newtheorem{prop}[thm]{Proposition}
\newtheorem*{lemma*}{Lemma}

\usepackage[f]{esvect}
\allowdisplaybreaks 

\journal{Elsevier}

\begin{document}

\begin{frontmatter}

\affiliation[ad6]{organization={Department of Industrial and Transport Engineering, Pontificia Universidad Católica de Chile}, city={Santiago}, 
country={Chile}}
\affiliation[ad5]{organization={Institute of Engineering Sciences, Universidad de O'Higgins}, city={Rancagua}, country={Chile}.}
\affiliation[ad1]{organization={Département d'Informatique, Université Libre de Bruxelles}, city={ Brussels}, country={Belgium}.}
\affiliation[ad2]{organization={Inria Lille-Nord Europe}, city={Villeneuve d'Ascq}, country={France}.}
\affiliation[ad4]{organization={Department of Electrical Engineering, Pontificia Universidad Católica de Chile and Instituto Sistemas Complejos de Ingeniería (ISCI)}, city={Santiago}, country={Chile}}
\affiliation[ad3]{organization={Department of Industrial Engineering, Universidad de Chile}, city={Santiago}, country={Chile}.}

 \author[ad6]{Pamela Bustamante-Faúndez\corref{cor1}\fnref{label2}}
 \ead{email address}
\author[ad5]{V\'ictor Bucarey L.}
\ead{victor.bucarey@uoh.cl}
\author[ad1,ad2]{Martine Labb\'e}
\ead{mlabbe@ulb.ac.be}
\author[ad4]{Vladimir Marianov}
\ead{marianov@ing.puc.cl}
\author[ad3]{Fernando Ordo\~nez}
\ead{fordon@dii.uchile.cl}

\title{Branch-and-price with novel cuts, and a new Stackelberg Security Game}

\begin{abstract}

Anticipating the strategies of potential attackers is crucial for protecting critical infrastructure. We can represent the challenge of the defenders of such infrastructure as a Stackelberg security game. The defender must decide how to allocate limited resources to protect specific targets, aiming to maximize their expected utility (such as minimizing the extent of damage) and considering that attackers will respond in a way that is most advantageous to them.

We present novel valid inequalities to find a Strong Stackelberg Equilibrium in both Stackelberg games and Stackelberg security games. We also \textcolor{black}{consider} a Stackelberg security game that aims to protect targets with a defined budget. We use branch-and-price in this game to show that our approach outperforms the standard formulation in the literature, and we conduct an extensive computational study to analyze the impact of various branch-and-price parameters on the performance of our method in different game settings. 

\end{abstract}

\begin{highlights}
    \item New valid inequalities for Stackelberg Games and Stackelberg Security Games.
    \item Using these valid inequalities and Branch-and-price, 
    \textcolor{black}{increases} the solution speed.
    \item New Stackelberg security game that protects targets at different costs, with a limited budget.
\end{highlights}

\begin{keyword}
Game theory \sep Stackelberg Games \sep Stackelberg Security Games \sep Optimization

\end{keyword}

\end{frontmatter}

\section{Introduction}

Security is a critical concern for governments, organizations, and individuals. Understanding and effectively managing security challenges involves analyzing strategic interactions between those defending against threats (defenders) and those posing the threats (attackers). Stackelberg games provide a valuable framework for addressing these security-related situations.  

Stackelberg security games (SSG) involve two players: a defender who first deploys a security strategy using limited resources and an attacker (or multiple attackers) who observes the defender's actions before attacking. The defender aims to maximize their payoff, taking into account the possible actions of the attacker, while the attacker seeks to optimize their own payoff based on the defender's initial moves. When there is no information about which potential attacker will make a move, these games are called Bayesian Stackelberg Games (Bayesian SG).

In SSGs, the defender implements an optimal mixed (or randomized) defense strategy. This mixed strategy represents a probability distribution over all the possible defense strategies. The attackers observe this and adjust their strategy for their best outcome. Due to its importance in security applications and being the dominant modeling choice in the literature \citep{bucarey2019coordinating,bucarey2019discussion,gan2014minimum,kar2017trends,kiekintveld2009computing, wilczynski2016stackelberg}, we find a strong Stackelberg equilibrium (SSE). In other words, when attackers have multiple optimal strategies, they select the most favorable for the defender.

A major challenge in Stackelberg games is the complexity of these problems. Solving Bayesian SGs and Bayesian SSGs is generally NP-Hard \citep{Conitzer2006_computing, LiConitzer2016}. Even in the simplest case, when the defender's strategy is to allocate resources to protect targets, enumerating all possible actions is intractable.

Several mixed-integer linear programming (MILP) formulations are available to address the NP-hardness challenge of finding a solution. 
We divide these approaches into two main categories. The first approach is based on noncompact formulations, which use variables corresponding to the defender's and attacker's strategies. The decision variables in this approach represent the probability of choosing each strategy \citep{Casorran2019, paruchuri2008playing}. The second methodology consists of modeling the problem using compact formulations, where the variables represent the frequency of defending each target 
\citep{kiekintveld2009computing, bustamante2023}, often referred to as ``marginal probabilities". 
Although the latter technique makes the problem more tractable, it may not always generate implementable defense strategies in practice \citep{bustamante2023, Korzhyk2010}.

When the solution of compact formulations is not implementable, it becomes necessary to use noncompact formulations. These models require an enumeration of all possible strategies through an 
\textcolor{black}{exponential} number of variables. To manage the complexity involved, we employ a branch-and-price technique. This approach, which has been previously utilized by \citet{Jain2010, Lagos2017} and \citet{Yang2013}, introduces variables as needed, improving efficiency. We introduce new general valid inequalities that are applicable to both Stackelberg games (SG) and Stackelberg security games (SSG). 
We demonstrate that these cuts are valid, and through computational experience, we show their effectiveness in reducing processing time.

Furthermore, we \textcolor{black}{consider} a game in which a defender has a limited budget to protect targets that can be attacked by an adversary. Each target has its own associated defense cost, which may depend on factors such as the distance needed to relocate patrol resources, the resources and infrastructure needed for defense, among others. This game frequently arises in real scenarios, yet there is no approach to solve it.

Our contributions are the following. First, we propose new 
valid inequalities for SGs and SSGs. Second, we  \textcolor{black}{consider} a SSG game whose objective is to protect targets with a defined total budget. Third, we use the branch-and-price method in this new game to show that our approach with the new valid inequalities outperforms the standard formulation in the literature as well as Benders' decomposition. Finally, we test different parameters for our branch-and-price approach and find the best configuration. 

The structure of this paper is as follows. In Section \ref{literatureReview}, we provide a review of the literature. We then introduce the formulations used throughout this paper and show new valid inequalities for SG and SSG in Sections \ref{notationAndFormulation} and \ref{SSG}, respectively.
 In Section \ref{methodology}, we propose a branch-and-price approach using our new valid inequalities, which we apply to address a new SSG problem. In Section \ref{Implementation}, we detail the branch-and-price implementation. We present experimental results in Section \ref{Experiments}, comparing our approach with state-of-the-art models.  We also explore the efficiency of the branch-and-price technique using different parameters. Finally, we state the conclusions in the Section \ref{Conclusion}.

\section{Literature Review}
\label{literatureReview}
The research on Stackelberg security games has experienced significant growth over the past decade. Several studies explore different aspects of these games, including mathematical formulations \cite{Casorran2019, kiekintveld2009computing, paruchuri2008playing}, solution approaches \cite{Jain2010, Lagos2017}, and real-world applications. Stackelberg security games are used in different types of security domains, ranging from transportation network security \cite{tsai2009iris} to border protection \cite{bucarey2019coordinating}, cybersecurity \cite{Zhang2021}, biodiversity protection in conservation areas \cite{FangFei2015}, and military defense \cite{kiekintveld2009computing}. These studies demonstrate the versatility and applicability of Stackelberg security games. 

Multiple mixed-integer linear programming (MILP) formulations address SGs and SSGs. \citet{paruchuri2008playing} introduced an MILP formulation, $(\text{DOBSS})$, to solve Bayesian SGs. This approach substitutes the follower's best response with linear inequalities, effectively transforming the leader's objective function into a linear form. \citet{kiekintveld2009computing} presented the compact formulation $(\text{ERASER})$ for SSGs, incorporating variables related to marginal probabilities. Although this formulation is more efficient, it does not always provide an implementable mixed defense strategy.
In addition, \citet{Casorran2019} extensively examined multiple formulations of Bayesian SSGs, exploring their relationships and characteristics. Their research introduced a new SSG formulation called $(\text{Mip-k-S})$, which describes the convex hull of feasible solutions (i.e., the perfect formulation) for scenarios involving a single attacker. Note that the letter k in $(\text{Mip-k-S})$ represents the number of attackers.

In noncompact formulations of Stackelberg games, the number of strategy-related variables can be intractable in realistic applications. An effective approach is the use of branch-and-price. For example, \citet{Jain2010} propose a generalized branch-and-price algorithm to solve a Stackelberg security game with a single attacker. They test their approach in a scenario where the defender has to create flight schedules for the Federal Air Marshals Service (FAMS), which involves protecting two targets per defense resource. They further test their approach for the case of protecting two to five targets per security resource.
The algorithm found optimal solutions for large instances using an efficient column generation approach that exploits a network flow representation and a branch-and-bound algorithm that generates bounds via a fast algorithm for solving security games with relaxed scheduling constraints.

In a different study by \citet{Lagos2017}, they introduce a branch-and-price-and-cut algorithm designed to address a Bayesian Stackelberg game scenario in which the defender has to allocate $m$ resources to protect individual targets (or nodes in a network) from possible attacks. Their algorithm is based on the formulation (Mip-k-G) proposed by \citet{Casorran2019} and incorporates a branching strategy that uses Lagrangian relaxation bounds, and stabilization techniques in the pricing problem. The study shows computational results specifically for a scenario involving one attacker.

\citet{Yang2013} introduce a branch-and-price technique for a generalized case of target protection considering both spatial and user-specified constraints and one attacker with bounded rationality. They discover that this approach is inefficient. Consequently, they propose an alternative method that is based on a noncompact formulation with relaxed constraints and a cutting-plane approach. 

Our research addresses a SSG problem consisting of protecting costly targets with a predetermined budget. 
Let us note that we can model the defender's strategy space as a Knapsack \textcolor{black}{constraint}. To the best of our knowledge, this problem has not been previously addressed in the SSG domain. A different problem that might be relevant is the Knapsack interdiction problem (KIP). In KIP, both the leader and the follower have their own knapsacks with capabilities, and both select items from a common pool. The leader chooses items to ensure the most unfavorable outcome for the follower. Multiple authors address this problem, such as \citet{fischetti2019, caprara2016bilevel, denegre2011interdiction}.

We extend the literature by introducing, first, new valid inequalities that strengthen the formulations of Bayesian Stackelberg games and Stackelberg security games. Second, we incorporate these constraints in a branch-and-price framework, achieving a solution speed that is twice as fast 
as the standard branch-and-price formulation in a new game. 
We demonstrate the versatility of our methodology by successfully applying it to a SSG problem focused on protecting specific targets given a specific budget.

\section{Stackelberg Games}
\label{notationAndFormulation}

In a Bayesian Stackelberg game, players try to optimize their payoffs in a sequential, one-off encounter.
In this model, the first player, called the \textit{leader}, has to face one of the $K$ other players or followers. Each follower acts or appears with a probability $p^k$. 

Let $I$ be the set of pure strategies for the leader, and $J$ be the set of pure strategies for the follower.  
If the leader chooses an action $i \in I$ and the follower $k$ performs an action $j \in J$, the first 
receives a payoff of $R_{ij}^k$ and the latter a payoff of $C_{ij}^k$.

A mixed strategy for the leader implies that he chooses a mix of pure strategies $i \in I$, each with a probability $x_i$. Similarly, a mixed strategy for the follower $k$ means that he chooses each pure strategy $j$ 
with a probability $q_j^k$. It is worth noting that, without loss of generality, we can assume that \textcolor{black}{the best response of the followers,} $\mathbf{q}=\{q_1^1,q_2^1,...,q_{|J|}^{|K|}\}$, is a pure strategy, as it constitutes the best response to a mixed strategy $\mathbf{x}=\{x_1,x_2,...,x_{|I|}\}$. In other words, $q_j^k \in \{0,1\}$ (we refer the reader to \citet{Casorran2019} for a proof).

The optimal solution in this \textcolor{black}{game} depends on how we define acting optimally or, in other words, the type of equilibrium we use. \citet{Breton1985SequentialSE} formalized the concepts of weak and strong Stackelberg equilibrium \citep{bard2013practical, dempe2002foundations}. In both cases, the leader chooses a strategy to maximize his utility, knowing that the follower will respond optimally. In a strong Stackelberg equilibrium, the follower favors the leader when multiple optimal strategies exist. While in a weak Stackelberg equilibrium, the follower chooses the option least beneficial to the leader in such cases. Every Stackelberg game has a strong Stackelberg equilibrium, but a weak Stackelberg equilibrium may not always exist. The strong Stackelberg equilibrium is the predominant choice in the literature \citep{gan2014minimum,kar2017trends,kiekintveld2009computing, wilczynski2016stackelberg}. We follow this trend, focusing on the strong Stackelberg equilibrium. 

In a Stackelberg game, the choice of the leader affects the choice of the follower. This structure gives rise to different formulations, including the following, denoted $(\text{D2})$ \citep{Casorran2019}:
\begin{align}
(\mbox{D2}) \nonumber\\
\max_{x,f,s,q} \,& \sum_{k \in K} p^k f^k \label{D2_mnodes1}\\
\mbox{s.t } 
&\textbf{x}^\intercal \textbf{1} = 1, \textbf{x} \ge 0 \label{D2_mnodes2}\\
&{\textbf{q}^k}^\intercal \textbf{1} = 1, \textbf{q}^k \in \{0,1\}^{|J|}  & \forall k \in K  \label{D2_mnodes3}\\
&f^k \le \sum_{i \in I} R_{ij}^k x_i + M(1-q_{j}^{k})  & \forall j \in J, k \in K \label{D2_mnodes4}\\
&s^k - \sum_{i \in I} C_{ij}^k x_i \le M(1-q_{j}^{k})  & \forall j \in J, k \in K \label{D2_mnodes5}\\
&0 \le s^k - \sum_{i \in I} C_{ij}^k x_i & \forall j \in J, k \in K \label{D2_mnodes5.1}\\
&s,f \in \mathbb{R}^k& \label{D2_mnodes6}
\end{align}

\noindent where the variables $f^k$ and $s^k$ represent the expected utility for the leader and follower of type $k$ when they face each other. The variables $x_i \in [0,1]$ represent the probability that the leader plays strategy $i \in I$, while $q_j^k \in \{0,1\}$ represents the probability of follower $k$ to select pure strategy $j$, and $M$ is a large number.

The objective function \eqref{D2_mnodes1} seeks to maximize the expected utility of the defender. Constraints \eqref{D2_mnodes2} express the mixed strategy employed by the leader. Equations \eqref{D2_mnodes3} specify that follower $k \in K$ selects the pure strategy $j \in J$. Constraints \eqref{D2_mnodes4}-\eqref{D2_mnodes5.1} ensure that both followers and the leader respond in an optimal manner. Finally, equations \eqref{D2_mnodes6} establish the nature of the variables $f$ and $s$.

This is a state-of-the-art formulation for SGs and SSGs, that facilitates the implementation of branch-and-price techniques, essential in games in which the number of available strategies tends to be intractable. Note that equations \eqref{D2_mnodes3} and \eqref{D2_mnodes4} use a big M parameter. We will take advantage of both of these elements in this article,
creating new and tighter valid inequalities, and using branch-and-price.

\citet{Casorran2019} show that the tightest correct $M$ values are:
\begin{itemize}
    \item In \eqref{D2_mnodes4}, $M=\max_{i \in I} \{\max_{l \in J} R_{il}^k-R_{ij}^k\} \; \forall j \in J, k \in K$.
    \item In \eqref{D2_mnodes5}, $M=\max_{i \in I} \{\max_{l \in J} C_{il}^k-C_{ij}^k\} \; \forall j \in J, k \in K$.
\end{itemize}

\newpage

\subsection{New valid inequalities for Stackelberg games}
\label{newcuts_SG}

This section introduces new valid inequalities for the $(\text{D2})$ Bayesian Stackelberg Game formulation. 

\begin{prop}
The following constraint is valid and dominates constraints \eqref{D2_mnodes4}:
\begin{equation}
f^k \leq \sum_{i \in I}R^k_{ij} x_i + \sum_{\substack{j' \in J: \\ j' \neq j}} \max_{i \in I}(R^k_{ij'} - R^k_{ij})q^k_{j'} \quad  \forall j \in J, \forall k \in K
\label{VI:1}
\end{equation}
\end{prop}

\begin{proof}
If $q^k_{j'} = 1$ for some $j' \in J$ and $k \in K$, then constraint 
\eqref{D2_mnodes4} becomes:
\begin{align*}
f^k &= \sum_{i \in I} R^k_{ij'} x_i \\
&=\sum_{i \in I}R^k_{ij} x_i + \sum_{i \in I}(R^k_{ij'}-R^k_{ij}) x_i \\ & \leq \sum_{i \in I}R^k_{ij} x_i + \max_{i \in I}(R^k_{ij'}-R^k_{ij}),\quad  \mbox{since} \sum_{i \in I} x_i = 1 \\
f^k &\leq \sum_{i \in I}R^k_{ij} x_i + \sum_{\substack{l \in J: \\ l \neq j}} \max_{i \in I}(R^k_{il} - R^k_{ij})q^k_{l} \quad  \forall j \in J, \forall k \in K, \mbox{ since }  
\sum_{j \in J} q^k_{j} = 1 \, \forall k \in K 
\end{align*}
which is the RHS of the inequality \eqref{VI:1}. Hence inequality \eqref{VI:1} is valid.

Further, after replacing the big M with the value given in \citet{Casorran2019}, constraint \eqref{D2_mnodes4} reads:

\begin{equation*} 
f^k \leq \sum_{i \in I}R^k_{ij} x_i + \max_{l \in J}\max_{i \in I}(R^k_{il} - R^k_{ij})(1-q^k_j) \;\;\;\; \forall j \in J, k \in K 
\end{equation*}
 which is clearly dominated by the above valid inequality \eqref{VI:1} since $1 - q^k_j = \sum_{\substack{j' \in J: \\ j' \neq j}} q^k_{j'}$.
 \end{proof}

We can also derive a similar inequality for the attacker objective function that dominates constraints \eqref{D2_mnodes5}:
\begin{equation}
s^k \leq \sum_{i \in I}C^k_{ij} x_i + \sum_{\substack{j' \in J: \\ j' \neq j}} \max_{i \in I} (C^k_{ij'} - C^k_{ij})q^k_{j'}  \; \;  \forall j \in J, \forall k \in K \label{VI:2}
\end{equation}

The following are valid inequalities that can improve the MILP relaxation, even though they do not dominate the original constraints \eqref{D2_mnodes4}-\eqref{D2_mnodes5}.
\begin{prop}
The following constraint is valid for each $i\in I$:

\begin{equation}
f^k \leq \sum_{j \in J}R^k_{ij} q^k_j  + \sum_{\substack{i' \in I: \\ i' \neq i}} \max_{j \in J} (R^k_{i'j} - R^k_{ij})x_{i'}
\label{sib:1}
\end{equation}

\end{prop}
\begin{proof}
 If $q^k_j = 1$, then constraint 
\eqref{D2_mnodes4} becomes:
 \begin{align*}
   f^k =& \sum_{i \in I} R^k_{ij} x_i \\
   =&\quad R^k_{ij} + \sum_{i' \in I} R^k_{i'j}x_{i'} -  R^k_{ij} \\
   =&\quad R^k_{ij} + \sum_{i' \in I} R^k_{i'j}x_{i'} -  \sum_{i' \in I}R^k_{ij}x_{i'},\quad \mbox{since} \sum_{i \in I} x_i = 1, \\
   =& \quad R^k_{ij} + \sum_{i' \in I} (R^k_{i'j} -  R^k_{ij})x_{i'}
   \\
   =& \quad R^k_{ij} + \sum_{\substack{i' \in I: \\ i' \neq i}} (R^k_{i'j} -  R^k_{ij})x_{i'}, \quad \mbox{since if $i'=i$, then the second term 
   is zero.}  \\
\leq & \quad R^k_{ij} + \sum_{\substack{i' \in I: \\ i' \neq i}} \max_{l \in J}(R^k_{i'l} -  R^k_{il})x_{i'} 
   \end{align*}

\textcolor{black}{Given that the second term of the RHS does not depend on the target $j$ that is attacked, we obtain:}
   \begin{align*}
   f^k \leq & \sum_{j \in J}R^k_{ij} q^k_j + \sum_{\substack{i' \in I: \\ i' \neq i}} \max_{j \in J}(R^k_{i'j} -  R^k_{ij})x_{i'}. \\
 \end{align*}
\end{proof}

 Again, this inequality can also be adapted to the follower objective, which would be: 
\begin{equation}
s^k \leq \sum_{j \in J}C^k_{ij} q^k_j  + \sum_{\substack{i' \in I: \\ i' \neq i}} \max_{j \in J} (C^k_{i'j} - C^k_{ij})x_{i'} \quad  \forall i \in I \label{sib:2}
\end{equation}

Let us note that constraints \eqref{sib:1}-\eqref{sib:2} are strengthening valid inequalities that do not replace the constraints \eqref{D2_mnodes4}-\eqref{D2_mnodes5}, nor \eqref{VI:1}-\eqref{VI:2}. 

The new constraints \eqref{VI:1} and \eqref{VI:2} lead to a new formulation, which we 
call $(\mbox{D2}_+)$ for SG:

\begin{align}
(\mbox{D2}_+) \nonumber\\
\max \,& \sum_{k \in K} p^k f^k\\
\mbox{s.t } 
&\eqref{D2_mnodes2},\eqref{D2_mnodes3}, \eqref{D2_mnodes5.1}-\eqref{VI:2} \nonumber
\end{align}

When we introduce the additional strengthening constraints \eqref{sib:1}-\eqref{sib:2} to the previous formulation, we 
denote this formulation as $(\text{D2}_+^{\prime})$ for SG.

\section{Stackelberg Security Games}
\label{SSG}

Let us consider a Bayesian Stackelberg security game where players aim to maximize their payoff in a sequential, one-off encounter. In this game, the defender faces a set of $k \in K$ attackers, each with a probability $p^k$ of acting or appearing. 

The defender has a set of strategies denoted as $I$. Every strategy within this set ($i \in I$) specifies \textcolor{black}{a} particular \textcolor{black}{subset of} targets protected by the security resources, which could vary from police personnel to canine units. \textcolor{black}{For instance,} depending on the game, it is feasible for a single security resource to cover several targets.  Let us note that we can view the leader's best \textcolor{black}{mixed strategy} as a combinatorial optimization problem over $I$. Therefore, the complexity of a security game is essentially determined by the set $I$ \citep{xu2016mysteries}.

On the other hand, the attackers have a strategy set $J$. We assume that each attacker targets only one objective. Thus, a pure strategy $j \in J$ corresponds to a single target $j$ under threat. 

Each player's profit depends only on whether the attacked target is protected. For every target $j$ within the set $J$ and for each type of attacker $k$ in set $K$, the defender's possible profits are $D^k(j|p)$ if the target is protected, and $D^k(j|u)$ if it is not. Similarly, the attacker $k$ gains payoffs of $A^k(j|p)$ and $A^k(j|u)$ based on the target $j$ being protected or unprotected, respectively. 
We assume that $D(j|p) \ge D(j|u)$ and $A(j|p) \le A(j|u)$.

\newpage
When addressing SSG, the $(\text{D2})$ formulation can be cast as:
\begin{align}
(\mbox{D2}) \nonumber \\
\max_{x,f,s,q} \,& \sum_{k \in K} p^k f^k \label{D2_1}\\
\mbox{s.t }
&f^k \le (D^k(j|p)-D^k(j|u)) \sum_{i \in I: j \in i} x_i  \nonumber \\ 
&+ D^k(j|u)  
+ (1-q_{j}^{k})M  & \;& \forall j \in J, k \in K \label{D2_2}\\
&s^k - (A^k(j|p)-A^k(j|u))\sum_{i \in I: j \in i} x_i \nonumber \\ 
&- A^k(j|u)  \le (1-q_{j}^{k})M   &  \; & \forall j \in J, k \in K \label{D2_3}\\
&0 \le s^k - (A^k(j|p)-A^k(j|u))\sum_{i \in I: j \in i} x_i  \nonumber \\ 
&- A^k(j|u)  &  \; & \forall j \in J, k \in K \label{D2_4}\\
&\sum_{i \in I} x_i = 1  & \;&   \label{D2_5}\\
&\sum_{j \in J} q_j^k = 1  & \;&  \forall k \in K \label{D2_6}\\
&q_j^k \in \{0,1\}& & \forall j \in J, k \in K \label{D2_7}\\
&s^k,f^k \in \mathbb{R} & & \forall k \in K& \label{D2_8}\\
& x_i \geq 0 & & \forall i \in I \label{D2_9}
\end{align}

In this context, the notation $j \in i$ denotes that the defender's strategy $i$ \textcolor{black}{protects} target $j$. \textcolor{black}{
Therefore, we associate each pure
strategy $i$ with a subset of targets $j \in J$ to be defended}. The objective function \eqref{D2_1}, aims to maximize the expected utility of the defender. Equations \eqref{D2_2} and \eqref{D2_3}-\eqref{D2_4} define that both the leader and followers must select strategies that optimize their respective expected profits, respectively. \textcolor{black}{For instance, if $q_j^k=1$ then $f^k = (D^k(j|p)-D^k(j|u)) \sum_{i \in I: j \in i} x_i + D^k(j|u)$}. Constraints \eqref{D2_5} express the mixed strategy of the defender. Additionally, the attacker can target only a single target, as stated in equation \eqref{D2_6}. The nature of the variables is defined in \eqref{D2_7}-\eqref{D2_9}. 

In the paper of \citet{Casorran2019}, they  \textcolor{black}{show} that the best values for $M$ are as follows:
\begin{itemize}
    \item In \eqref{D2_2}, $M=\max_{l \in J} \{D^k(l|p),D^k(l|u)\} - \min \{D^k(j|p),D^k(j|u)\} \; \forall j \in J, k \in K$.
    \item In \eqref{D2_3}, $M=\max_{l \in J} \{A^k(l|p),A^k(l|u)\} - \min \{A^k(j|p),A^k(j|u)\} \; \forall j \in J, k \in K$.
\end{itemize}

\subsection{New valid inequalities for Stackelberg security games}
\label{newcuts_SSG}
This section introduces novel valid inequalities for Bayesian Stackelberg Security Games. These constraints are equivalent to those we derived in Section \ref{newcuts_SG}, but in the context of Stackelberg security games.

\begin{prop}
The following constraint is valid and dominates constraints \eqref{D2_2}:
\begin{align}
f^k \leq  & (D^k(j|p) - D^k(j|u)) \sum_{\substack{i \in I: \\ j \in i}} x_i + D^k(j|u) \nonumber \\ 
&+ \sum_{\substack{j' \in J: \\ j' \neq j}} (D^k(j'|p) - D^k(j|u)) q^k_{j'}  \; \; \;\; \forall j \in J, k \in K
\label{VI_SSG:1}
\end{align}

\end{prop}

Analogously, we can 
derive a similar SSG inequality for the attacker's utility function. The following dominates  constraints \eqref{D2_3}: 
\begin{align}
 s^k \leq \; & (A^k(j|p) - A^k(j|u)) \sum_{\substack{i \in I: \\ j \in i}} x_i  + A^k(j|u) \nonumber\\ 
& +\sum_{\substack{j' \in J: \\ j' \neq j}} (A^k(j'|u) - A^k(j|p)) q^k_{j'} \; \; \forall j \in J, \forall k \in K
\label{VI_SSG:2}
\end{align}  

Let us note that these constraints \eqref{VI_SSG:1}-\eqref{VI_SSG:2} are equivalent to \eqref{VI:1}-\eqref{VI:2} for SSG.

\begin{prop}
The following constraints are valid:
\begin{align}
&f^k \leq \sum_{j \in J} D^k(j|p) q^k_j \label{SSG_str_1} \\
&s^k \leq \sum_{j \in J} A^k(j|u) q^k_j \label{SSG_str_2}
\end{align}
\end{prop}

These constraints are equivalent to \eqref{sib:1}-\eqref{sib:2} for SSG. \noindent For a detailed proof of constraint \eqref{SSG_str_1}, see Appendix \ref{proof_ssg}.

We call the new formulation with the \textcolor{black}{dominating} constraints \eqref{VI_SSG:1}-\eqref{VI_SSG:2} as $(\mbox{D2}_+)$ for SSG:
\begin{align}
(\mbox{D2}_+) \nonumber\\
\max \,& \sum_{k \in K} p^k f^k\\
\mbox{s.t } 
&\eqref{D2_4}-
\eqref{VI_SSG:2} \nonumber
\end{align}

If we add the strengthening constraints \eqref{SSG_str_1}-\eqref{SSG_str_2} to the previous formulation, we call this formulation $(\text{D2}_+^{\prime})$ for SSG.

Note that for the compact SSG formulation $(\text{ERASER})$ \citep{kiekintveld2009computing, bustamante2023}, the constraints \eqref{VI_SSG:1} and \eqref{VI_SSG:2} are also valid but need the transformation:
 $\sum_{\substack{i \in I: j \in i}} x_i  = c_j \; \forall j \in J$. It is also possible to add constraints \eqref{SSG_str_1}-\eqref{SSG_str_2}, resulting in $(\text{ERASER}_+^{\prime})$.

\section{The budget constrained SSG and its solution method} 
\label{methodology}

To test our new valid inequalities, we use a Stackelberg Security Game in which each target has a defense cost, and there is a limited budget available for defense. In this game, each pure strategy of the defender corresponds to protecting a subset of targets in such a way that the budget is not exceeded.  The defender selects the best mixed strategy, understanding that the follower will take this decision into account when deciding his optimal move.

Let us note that compact formulations such as (ERASER) or (Mip-k-S) do not guarantee a solution of this problem, since the solutions obtained may not be implementable in practice.
\textcolor{black}{A set of constraints is implementable if any random mixed strategy under that set is a convex
combination of pure strategies that satisfy the constraints \citep{budish2013designing}.}

The optimal solution of a compact formulation is only implementable when the defender space meets certain requirements. For example, \citet{bustamante2023} focus on a category of problems in which the \textcolor{black}{constraints} representing the defender's strategies is a perfect formulation, \textcolor{black}{i.e. they describe the convex hull of the incidence vectors of the pure strategies}. When a linear integer optimization problem has a perfect formulation, we can solve it as a continuous linear programming problem for that perfect formulation.
They show that solutions derived from a compact formulation of such problems are always implementable. Furthermore, they establish that upon finding a solution to a compact formulation, the corresponding mixed strategy can be found efficiently in polynomial time through column generation, provided that the defender strategy space has a polynomial number of constraints or an exponential number of constraints, separable in polynomial time. In the context of the game we study, however, the implementability of solutions using compact formulations cannot be guaranteed.
Therefore we use a noncompact formulation, with a branch-and-price approach that avoids full enumeration of the strategies.

We work with the non-compact formulation (D2), as it \textcolor{black}{is convenient for using} branch-and-price, unlike (DOBBS) and (Mip-k-G), which require branch-and-price-and-cut tailored for the game \citep{Lagos2017}.

\subsection{Column generation master problem}

Let us consider the new SSG formulation $(\text{D2}_+)$. 
To apply branch-and-price, we need a structure that easily allows the addition of new variables. We use the following transformation, proposed by \citet{Jain2010}:
\begin{align}
\sum_{i \in I: j \in i} x_i = \sum_{i \in I} x_i P^i_j
\end{align}

\noindent The set $\mathcal{P}$ contains all binary vectors $\bm{P}^i$ that encode pure defense strategies, i.e., $P^i_j = 1$ if $j$ is protected in strategy $i$, and $P^i_j = 0$ otherwise. Then, we can rewrite our new SSG formulation for the master problem $(\text{D2}_+)$ as: 
\begin{align}
(\mbox{D2}_+) \nonumber \\
\max_{x,f,s,q} \,& \sum_{k \in K} p^k f^k \\
\mbox{s.t } & \eqref{D2_5}-\eqref{D2_9} \nonumber \\
&f^k -  (D^k(j|p) - D^k(j|u)) \sum_{i\in I} P_j^{i} x_i  \nonumber\\ 
&- \sum_{\substack{j' \in J: \\ j' \neq j}} (D^k(j'|p) - D^k(j|u)) q^k_{j'} \leq D^k(j|u) \; &(w_j^k)\; & \forall j \in J, k \in K \\
&s^k -  (A^k(j|p) - A^k(j|u)) \sum_{i\in I} P_j^{i} x_i \nonumber \\ 
&- \sum_{\substack{j' \in J: \\ j' \neq j}} (A^k(j'|u) - A^k(j|p)) q^k_{j'}  \leq A^k(j|u) &(y_j^k) \; & \forall j \in J, k \in K \\
&- s^k + (A^k(j|p) - A^k(j|u)) \sum_{i \in I}P^i_{j} x_i \le - A^k(j|u) \; &(z_j^k)\; & \forall j \in V, k \in K \\ 
&\sum_{i \in I} x_i = 1 \; &(h) \;& 
\end{align}

The definition of the set $\mathcal{P}$ varies depending on the game studied, reflecting what the defender intends to protect. In our new game, each target $j$ has a defense cost of $w_j$, and a total budget of $W$ is available for defense. In this situation, we define the set $\mathcal{P}$ as:
\begin{align}
\mathcal{P} = \{& P \in \{0,1\}^{\vert V \vert}: \sum_{j \in V} w_j P_j \le W \} \label{eq:knapsack}
\end{align}

Let us note that \eqref{eq:knapsack} exhibits the characteristic structure of the well-known Knapsack Problem. Our restricted master problem (or RMP for short) starts including a \textcolor{black}{subset} of feasible columns $P \in \mathcal{P'}$, and we add new columns through a column generation approach. 

The parentheses on the right in the formulation  $(\text{D2}_+)$ represent dual variables associated 
with every constraint.

\subsection{The pricing problem}
During branch-and-price, we use the reduced cost of $x$ as part of our pricing problem to find new columns $P$ to add.  Recall that given a system $\{ \min c^T x: Ax \le b, x \ge 0 \}$, we can compute the reduced cost vector of $x$ as $c - A^T y$, where $y$ is the dual cost vector. 

\textcolor{black}{Given optimal primal and dual solutions to the LP relaxation of the RMP}, the reduced cost of $x_i \;\, \mbox{s.t } \, P_i \in \mathcal{P}$ is:
\begin{align}
cost_i= &- h + \sum_{j \in J}P_j \sum_{k \in K} (w_j^k (D^k(j|p) - D^k(j|u))+y_j^k (A^k(j|p) - A^k(j|u)) \nonumber \\ 
&- z_j^k (A^k(j|p) - A^k(j|u)))  \label{red_cost_xi}
\end{align}

Since we generally generate one variable $x_i$ per iteration, we drop the index $i$ of $P_j^i$. 
If the reduced cost is strictly positive, we need to add columns. Otherwise, we encountered the optimal solution.
In order to find new columns (if needed), the pricing problem is:
\begin{align}
\zeta =\max \{&  - h + \sum_{j \in J} P_j \sum_{k \in K}(w_j^k (D^k(j|p) - D^k(j|u))+y_j^k (A^k(j|p) - A^k(j|u)) \nonumber\\ 
&- z_j^k (A^k(j|p) - A^k(j|u))): P \in \mathcal{P} \} \label{eq:generalized_pricing}
\end{align}

\noindent The expression \eqref{eq:generalized_pricing} allows us to generate columns in the branch-and-price according to the problem we are addressing. 
In our methodology, we tested the GRASP \citep{deleplanque2020} and the Ratio Greedy algorithm to address this pricing problem. Additionally, we used a solver when we could not find \textcolor{black}{heuristically} new columns to add.

It is important to note that whether we are utilizing formulations $(\text{D2})$, $(\text{D2}_+)$, or $(\text{D2}_+^{\prime})$, the pricing problem in equation \eqref{eq:generalized_pricing} remains unchanged. This is because the pricing problem relies on the reduced cost of variables $\bm{x}$, which remains unaltered by 
introducing our new constraints.

\subsubsection{Farkas pricing }

In each iteration of the column generation scheme, we introduce a new variable $x_i$ with a positive reduced cost (if any). If there are no new columns to add, it means that we found an optimal LP solution, prompting us to initiate branching on binary variables $q_j^k$. However, some branches might appear as infeasible only because the number of columns added so far is insufficient. In fact, during the experiments, we realized that if there are enough columns, this problem does not arise. For an illustrative example, refer to Appendix \ref{infeasibility}.

\textcolor{black}{To address this issue}, we use Farkas pricing to detect if the infeasibility found at a branch is real or due to an insufficient number of columns added so far. Recall that by Farkas’ Lemma, a system of linear inequalities $Ax \le b$ is infeasible if and only if the system $y^T A=0, y^T b < 0, y\ge 0$ is feasible. This means that if a system is not feasible, then there exists a vector $\rho$ with $\rho A=0$ , $\rho b<0$ , $\rho \ge0$ , which we can interpret as a ray in the dual, in the direction of which the objective function decreases (if the RMP is maximization) or increases (if RMP is minimization) without limit.

The idea of the Farkas pricing problem is to turn the unfeasible branch into a feasible one, destroying this proof of infeasibility. 
To do so, we need to add a variable x to the MP whose coefficient column a(x) in the matrix A is such that $\rho a(x)<0$. For a maximization MP, we can find such variable by solving: $\min {\rho a(x)}=\max {-\rho a(x)}$, which is the standard pricing problem with null cost coefficients for the objective function of MP and the dual ray vector instead of the dual variable vector.

\section{Implementation}
\label{Implementation}

In this section, we explain the configuration of our branch-and-price algorithm. Our study involves varying the number of initial columns, 
determining the number of columns generated during the pricing phase, using heuristic pricing methods, and implementing stabilization techniques. Further details are provided below.

\subsection{Initial columns}

Before we solve the problem using the branch-and-price method, it is important to first establish a pool of initial \textcolor{black}{columns}. How well we choose these starting variables can significantly impact the time it takes to find a solution. We experiment with generating different numbers of columns using the GRASP algorithm \citep{deleplanque2020}. We chose this heuristic as it easily allows the generation of multiple columns, unlike the classical Ratio Greedy algorithm which only generates one column.  In order to 
assign a benefit of protecting each target $j$ for the heuristic, we use a proxy of the gradient of the objective function with respect to $x_i$, i.e., $\sum_{k \in K} p^k(D^k(j|p)-D^k(j|u))$.

\subsection{Heuristic pricing problem}

We conducted initial tests to address the pricing problem using two methods: the GRASP algorithm \citep{deleplanque2020} and the Ratio Greedy algorithm. Our findings indicate that contrary to what happens with the initial columns, the Ratio Greedy approach outperformed the GRASP algorithm in terms of speed. Therefore, we use the Ratio Greedy heuristic to solve the pricing problem faster (in an approximate manner). 

If the heuristic does not find a new column to add, we will solve the problem as a MILP with a solver. This procedure allows us to explore more nodes in the branch-and-bound tree and find better solutions within the time limit, compared to just using the exact method that may take longer. Using an heuristic pricer involves adding variables that do not have the lowest reduced cost during the initial iterations, which fastens convergence \cite{Blanco2023,pessoa2010}. It is worth mentioning that heuristic pricers additionally help in preventing degeneracy \citep{Benati2022, Blanco2021}.

\subsection{Stabilization for {\normalfont $(\text{D2})$}}
We follow the stabilization proposed by 
\citet{Lagos2017} and \citet{pessoa2013}. In this approach, we use a vector of dual variables defined as the convex combination of the current solution of the dual problem and the previous vector of dual variables. With stabilization, we initially add variables that do not have the lowest reduced cost, which accelerates the convergence \cite{pessoa2010}.

Let $\pi_t=(w,y,z,h)$ be a ``stabilized" vector of dual multipliers in iteration $t$, $\pi_{RMP}$ be the current dual solution of the restricted master problem, and $\Delta \in [0,1]$ be the 
weight of the current dual solution in vector $\pi_t$.

\begin{algorithm}
\caption{Stabilization}\label{algorithm-label}
\begin{algorithmic}[1]
\State $k \gets 1, \pi_0 \gets \pi_{RMP}$, $\Delta$ \textcolor{black}{is a fixed value}
\State $\tilde{\Delta} \gets [1 - k \cdot (1 - \Delta)]_+$
\State $\pi_{t} \gets \tilde{\Delta} \pi_0 + (1 - \tilde{\Delta}) \pi_{t-1}$
\State $k \gets k + 1$
\State Call the pricing oracle on $\pi_{t}$
\While{$cost(P,\pi_{RMP}) < 0$ }
    \State Go to Step 2
\EndWhile
\State $t \gets t + 1$, solve the master problem and go to Step 0
\end{algorithmic}
\end{algorithm}

  Algorithm 1 starts by initializing  variables, including the weight $\Delta$, a vector of dual variables $\pi_0$, and iterator $k$. 
   It then iteratively solves two key problems: the Restricted Master Problem (RMP) and the pricing problem. The algorithm continues to iterate as long as the reduced cost of variables $x$ is greater than 0. 
  
  We use a dual vector $\pi_t$ during each iteration \textcolor{black}{$t$} to solve the pricing problem. This vector is based on a weighted combination of the previous $\pi_{t-1}$ values and the $\pi_{RMP}$ dual values obtained from the RMP solution. The pricing problem finds a column $P$. If the reduced cost evaluated on the column $P$ and using the dual values $\pi_{RMP}$ is negative \textcolor{black}{ (i.e. $cost(P,\pi_{RMP})<0$)}, the algorithm will go to Step 2.
  After this, the algorithm adds variables $x_i$ to the problem if needed. 
   The algorithm 
  continues until the reduced cost condition is no longer positive.

\section{Computational Experiments}
\label{Experiments}

We compare our proposed formulations $(\text{D2}_+)$ and $(\text{D2}_+^{\prime})$, with the original formulation $(\text{D2})$, using branch-and-price. We run different tests using our approach, comparing different parameters and variants of the formulations.
 
We denote the set of targets as $J$, with $|J|=n$, and $K$ for the set of attackers, with $|K|=k$. We use penalty matrices with values randomly generated between $0$ and $5$, and reward matrices' values randomly generated between $5$ and $10$.

We employ the following metrics to summarize our results: the total time to solve the integer problem, the time to solve the linear relaxation of the mixed integer problem, the total number of nodes explored in the branch-and-bound (B\&B) tree, the percentage of optimality gap at the root node, the peak memory usage during computation, and the number of generated columns. Let us note that the number of nodes represents the number of processed nodes in all runs, including the focus node. 

We performed our experiments in an 11th Gen Intel Core i9-11900, 2.50GHz, equipped with 32 gigabytes of RAM, 16 cores, 2 threads per core, 
and running the Ubuntu operating system release 20.04.6 LTS. We coded the experiments in Python v.3.8.10 and SCIP v.8.0.0 as the optimization solver \citep{BestuzhevaEtal2021OO}, considering a 1-hour solution time limit.

\subsection{Budget constraints}

For the problem 
of protecting targets including budget constraints, we consider $J = \{20,30,40\}$ and $K = \{2,4,6\}$. 
For every configuration, we consider five different instances. 

We create instances of the knapsack problem, which include both costs and a budget, following the guidelines set by \citet{Kellerer2004}. We specifically concentrate on uncorrelated instances, selecting the cost $w_j$ of protecting a target randomly from the range $[1,100]$. 
We determine the budget for each instance as $B = \frac{h}{H+1} \sum_{j=1}^n w_j$, where $h = 1, \dotsc ,H$ represents the number of the test instance. In order to ensure fractional solutions and prevent the defender's strategy space from being a perfect formulation (which would allow for solving via a compact formulation), we adjust the budget by adding a fractional value within $[0.05,0.95]$.

\subsubsection{Comparison of different numbers of Initial Columns using GRASP heuristic and solver.} 

We study the impact of the number of initial columns within the branch-and-price framework. We conduct experiments employing different number of initial columns, specifically $\{1, 40, 80\}$, considering both the original $(\text{D2})$ formulation and our proposed $(\text{D2}_+)$ formulation.

\begin{figure}[H] 
    \centering 
    \includegraphics[width=0.9\textwidth]{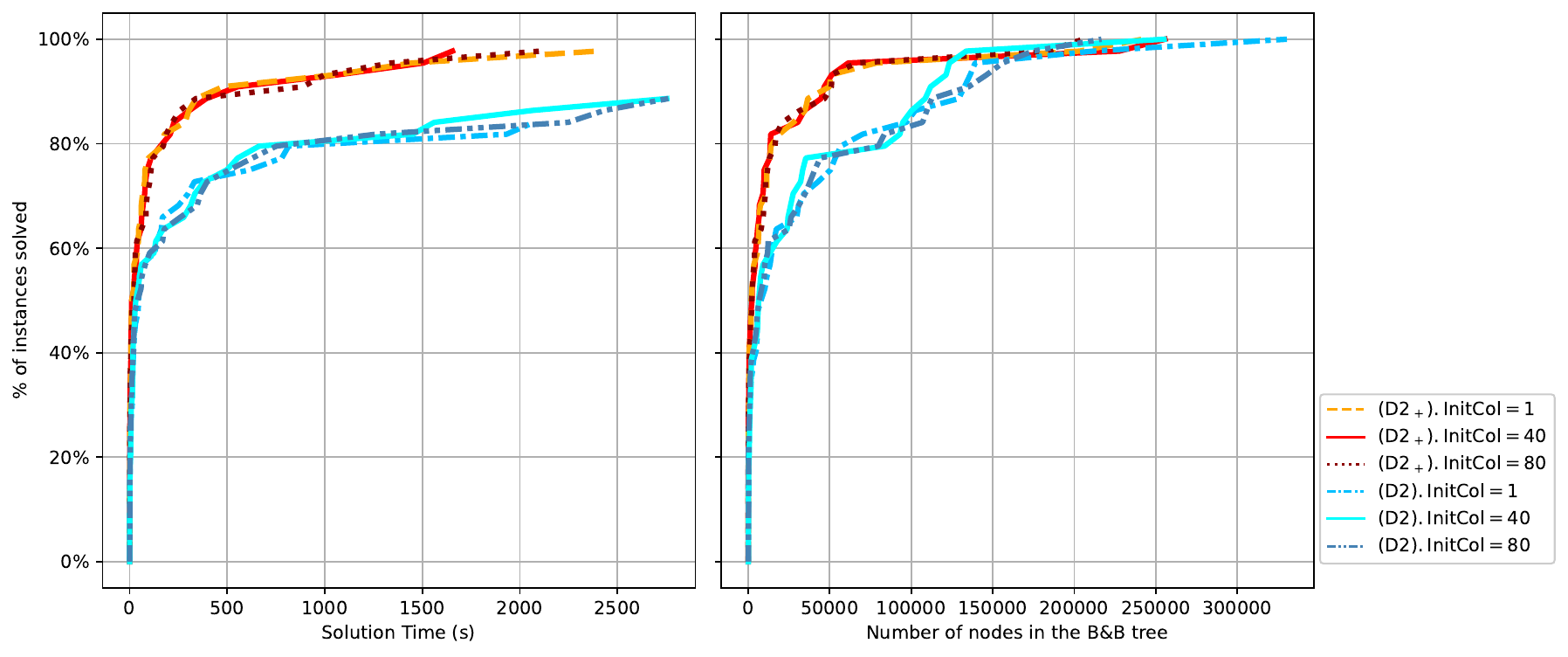}
    \caption{Time to solve the integer problem, and number of nodes in the B\&B tree. $ \text{InitialCols}=\{1, 40, 80\}$}\label{fig:1}
\end{figure}

From our experiments, we conclude that using a single column in the $(\text{D2}_+)$ and {$(\text{D2})$} formulations results in a slightly slower solving process.  
Using either 40 or 80 initial columns with (D2) does not yield significant differences. 
Our observations indicate that the $(\text{D2})$ formulation experiences more branch-and-bound nodes than $(\text{D2}_+)$. This is related to 
a longer solution time, as shown in Figure \ref{fig:1}. 
Also, the number of chosen columns in the optimal solution is always less or equal to $|J|$. Therefore, 
we opted to use $|J|$ initial columns for the following experiments, unless stated otherwise. 

\textcolor{black}{Let us note that when considering only the instances solved within the time limit, $(\text{D2}_+)$ generates, on average, only $52\%$ of the total number of columns that $(\text{D2})$ does.} Also, there is a clear difference in solution times between $(\text{D2})$ and $(\text{D2}_+)$. \textcolor{black}{When only considering instances solved within the time limit, $(\text{D2}_+)$ uses 19\% of the time $(\text{D2})$ requires, on average for all variants (1, 40, and 80 initial columns). Additionally, $(\text{D2}_+)$ solves each of the 80\% of the instances in under 170 seconds, on average for all variants. In contrast, $(\text{D2})$ needs an average of 750 seconds to solve the same percentage of instances. 
When examining 90\% of the instances, $(\text{D2})$ can not solve all of the instances within the time limit, whereas $(\text{D2}_+)$ solves them in 640 seconds on average.}

\begin{figure}[H] 
    \centering 
    \includegraphics[width=0.9\textwidth]{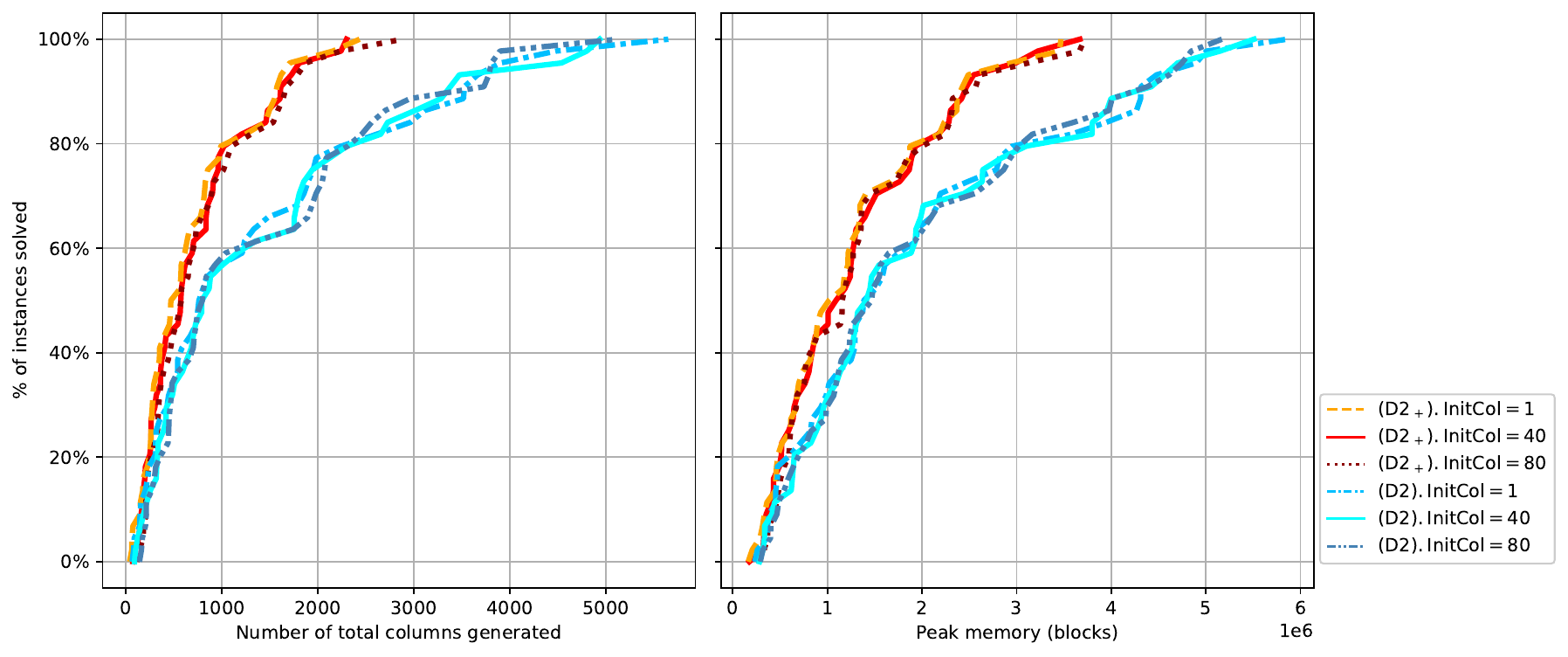}
    \caption{Number of total columns generated, and peak memory used to solve the integer problem. $ \text{InitialCols}=\{1, 40, 80\}$}\label{fig:2}
\end{figure}

Also, as 
expected, there is a direct correlation between the number of columns generated and the peak memory used (see Figure \ref{fig:2}). 

\subsection{Comparison of the number of columns generated during the pricing problem, using Ratio greedy algorithm and solver} 

We study how the number of columns we add to the master problem in each 
branch-and-price iteration affects the performance. We run tests with different numbers of columns, specifically [1,5,10], and we compare the results for both the original $(\text{D2})$ model and our improved $(\text{D2}_+)$  model.
Recall that in contrast to the stage where we generate initial columns through the GRASP algorithm, at this stage, we utilize the Ratio Greedy algorithm.

\begin{figure}[H] 
    \centering 
    \includegraphics[width=0.9\textwidth]{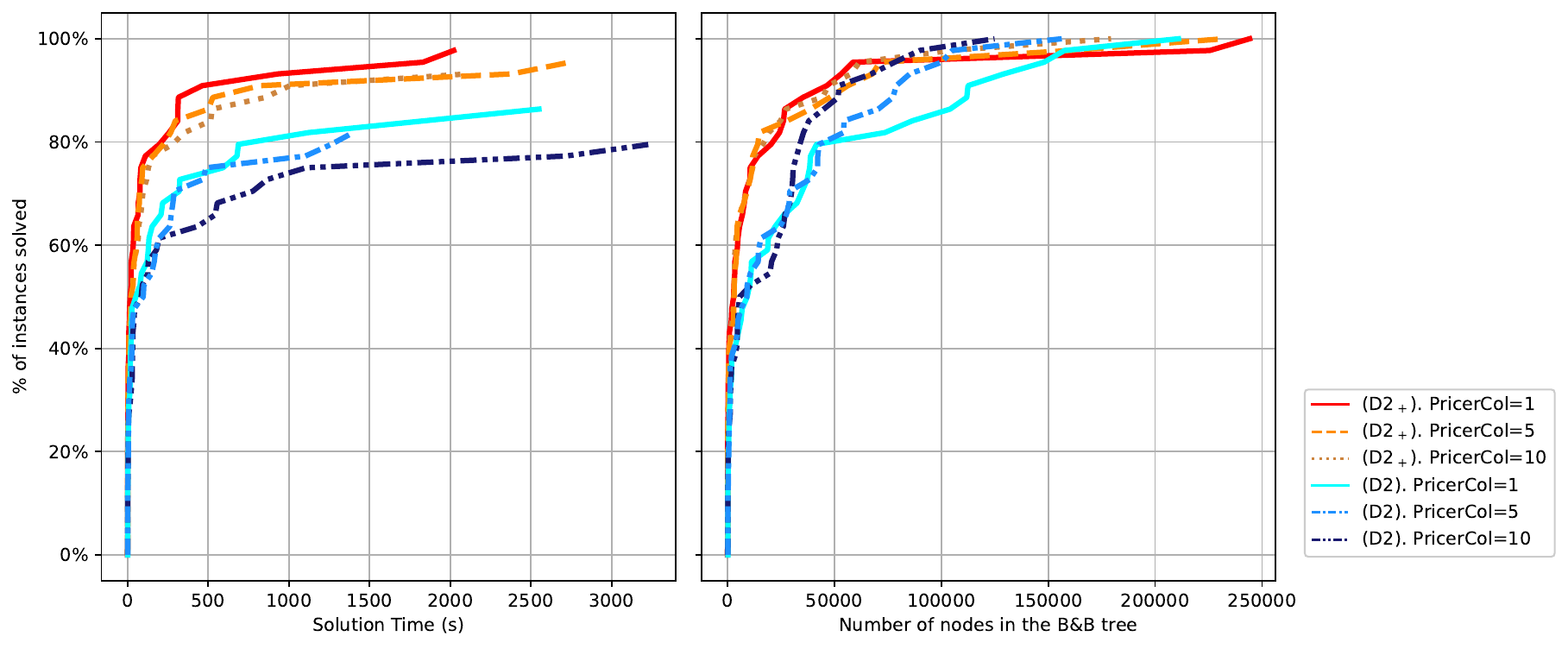}
    \caption{Time to solve the integer problem, and number of nodes in the B\&B tree, considering different numbers of columns in the pricing problem.
    }\label{fig:6}
\end{figure}

\begin{figure}[H] 
    \centering 
    \includegraphics[width=0.9\textwidth]{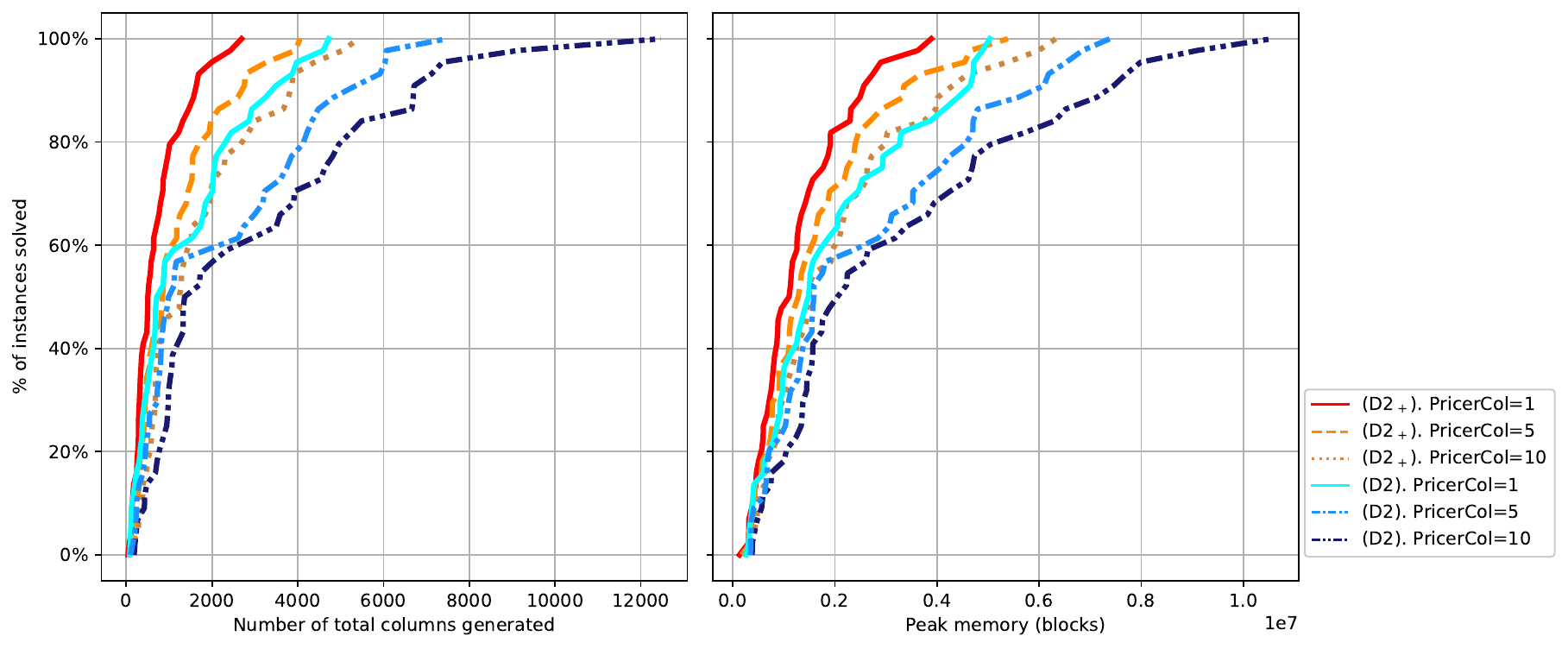}
    \caption{Number of total columns generated, and peak memory used to solve the integer problem, considering different numbers of columns in the pricing problem.}\label{fig:7}
\end{figure}

We conclude that generating one column per pricing problem iteration is the most efficient for our problem (Figure \ref{fig:6}). This means we only add one new variable to the master problem during each iteration. This can have some advantages. It can simplify the pricing problem by requiring us to identify the best variable rather than searching for a set of variables that meet specific criteria. 
Furthermore, this technique prevents the generation of unnecessary columns (as illustrated in Figure \ref{fig:7}), which could otherwise increase the master problem's size and memory usage and complicate the solving process. In our subsequent experiments, we adhere to this practice of generating one column per iteration of the pricer.

\subsubsection{Impact of constraints: {\normalfont $(\text{D2})$} vs. {\normalfont $(\text{D2}_+)$ (defender, attacker, both)} vs. {\normalfont $(\text{D2}_+^{\prime})$ (defender, attacker, both)}}

We now examine the new valid inequalities' impact on the efficiency of the formulation. We study the $(\text{D2}_+)$ formulation focusing on three variants: one that uses exclusively the new valid inequalities for the attacker, denoted as $(\text{D2}_{+att})$, another that uses exclusively the valid inequalities for the defender, $(\text{D2}_{+def})$, and a third that combines both $(\text{D2}_+)$. 

Besides this, we also conducted experiments considering the strengthening valid inequalities, using the valid inequalities just for the attacker $(\text{D2}_{+}^{\prime att})$, just for the defender 
$(\text{D2}_{+}^{\prime def})$, and for both 
$(\text{D2}_+^{\prime})$. We compare these results with the original $(\text{D2})$  formulation. In our instances,
$(\text{D2}_+^{\prime})$ is, on average, 72\% faster than $(\text{D2})$.

\begin{figure}[H] 
    \centering 
    \includegraphics[width=0.9\textwidth]{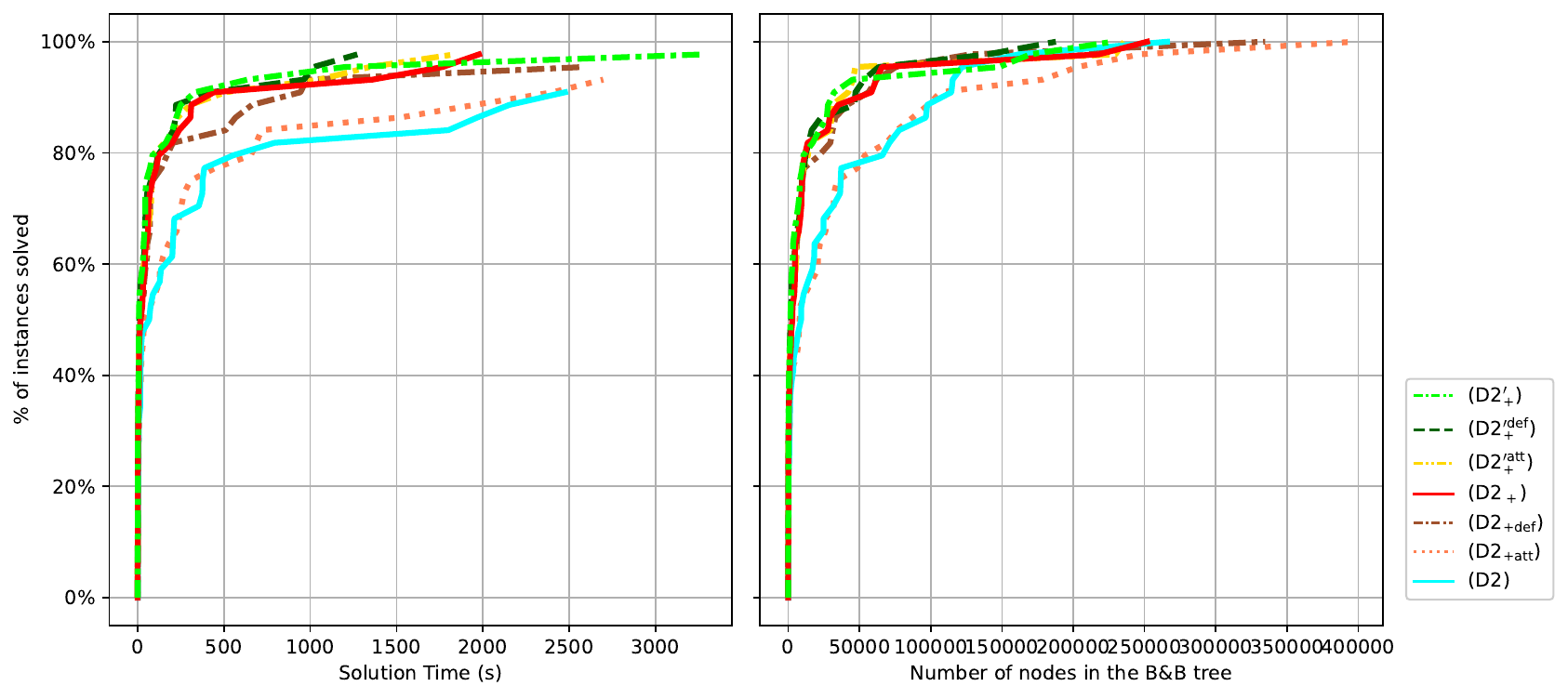}
    \caption{
    Time to solve the integer problem, and number of nodes in the B\&B tree, considering new and original constraints.
    }\label{fig:3}
\end{figure}

Figure \ref{fig:3} illustrates our results. Our findings indicate that the combined valid inequalities $(\text{D2}_+)$ yield the most favorable results compared to just using valid inequalities for the attacker $(\text{D2}_{+att})$, or defender $(\text{D2}_{+def})$. 
Additionally, it is worth noting that the valid inequalities for the defender have a 
more significant impact on performance than those used for the attacker.

Also, the $(\text{D2}_+)$ formulation requires less than 
\textcolor{black}{$450$} seconds to solve each of the $90\%$ of the instances, whereas solving each of the instances of the same percentage of instances with $(\text{D2})$ takes 
less than 
\textcolor{black}{2480 seconds}. Although $(\text{D2})$ initially performs well, especially with smaller instances, its efficiency diminishes as the instances increase in size. Among the examined $(\text{D2}^{\prime}_+)$ formulations variants, the slowest performance is observed in $(\text{D2}^{\prime att}_{+})$, and the fastest in $(\text{D2}^{\prime def}_{+})$. We consider that both $(\text{D2}^{\prime def}_{+})$ are $(\text{D2}^{\prime}_+)$ are competitive, as they solve almost the same amount of instances in the time limit. 
Our $(\text{D2}_{+})$ formulation can solve almost all instances \textcolor{black}{(98\%)} within an hour. 
In contrast, the state-of-the-art $(\text{D2})$ can only handle approximately 90\% of instances in the same time frame. \textcolor{black}{When only considering instances solved within the time limit, $(\text{D2}_{+})$ requires only 18.7\% of the time needed by $(\text{D2})$.}  

Note that we have opted not to use Farkas pricing in this subsection due to its tendency to slow the solving process. Rather, we are incorporating an additional $|J|$ initial columns to not need to use Farkas pricer.

\begin{figure}[H] 
    \centering 
\includegraphics[width=0.9\textwidth]{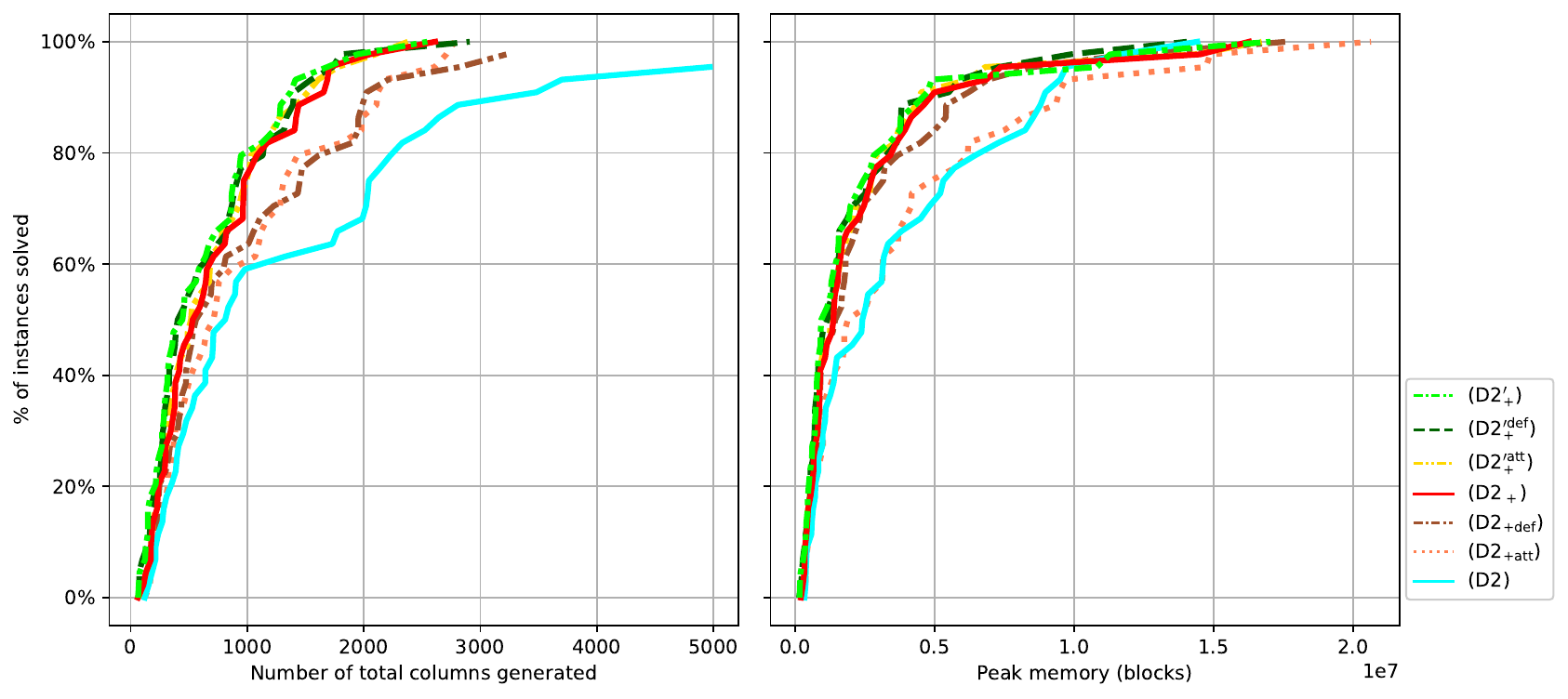}
    \caption{
    Number of total columns generated, and peak memory used to solve the integer problem, considering new and original constraints.
    }\label{fig:4}
\end{figure}

\begin{figure}[H] 
    \centering 
    \includegraphics[width=0.50\textwidth]{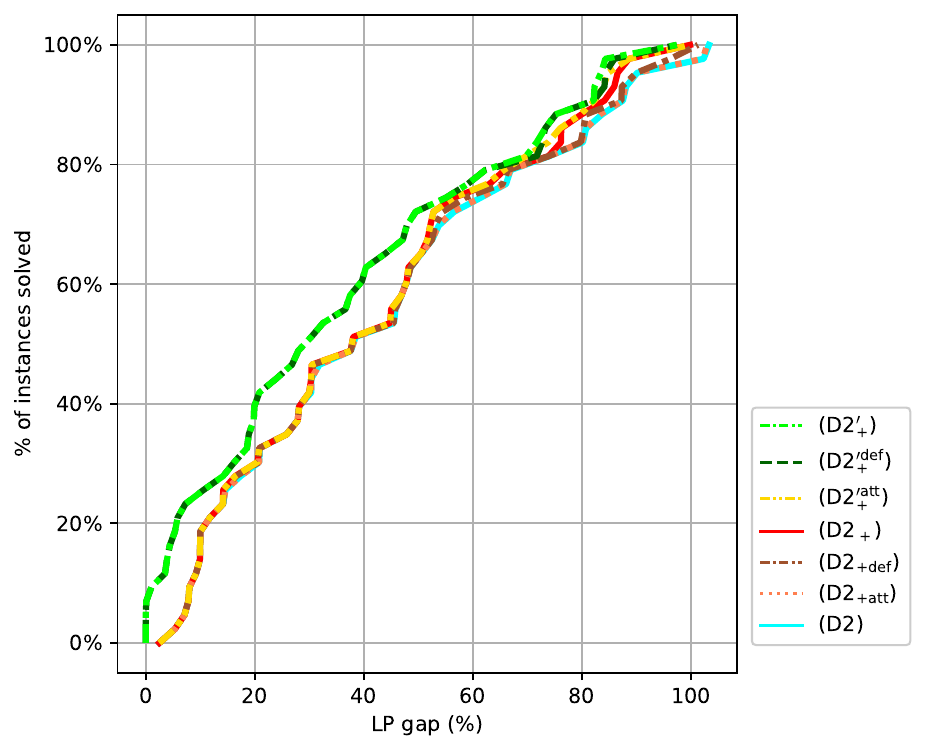}
    \caption{LP gap, considering new and original constraints.}\label{fig:5}
\end{figure}

On average, the total number of columns generated in the $(\text{D2}_+)$ formulation \textcolor{black}{is 
52\% of the number of columns generated} in the formulation $(\text{D2})$. Furthermore, as in the previous experiment, there exists a direct correlation between the quantity of generated columns and the peak memory usage (refer to Figure \ref{fig:4}). 
Additionally, it is important to highlight that the percentage of LP Gap for the $(\text{D2}_{+}^{\prime})$ variants is smaller than for the $(\text{D2}_{+})$ formulation (Figure \ref{fig:5}).

\subsection{Comparison of stabilization vs. no stabilization.}

\begin{figure}[H] 
    \centering 
    \includegraphics[width=0.9\textwidth]{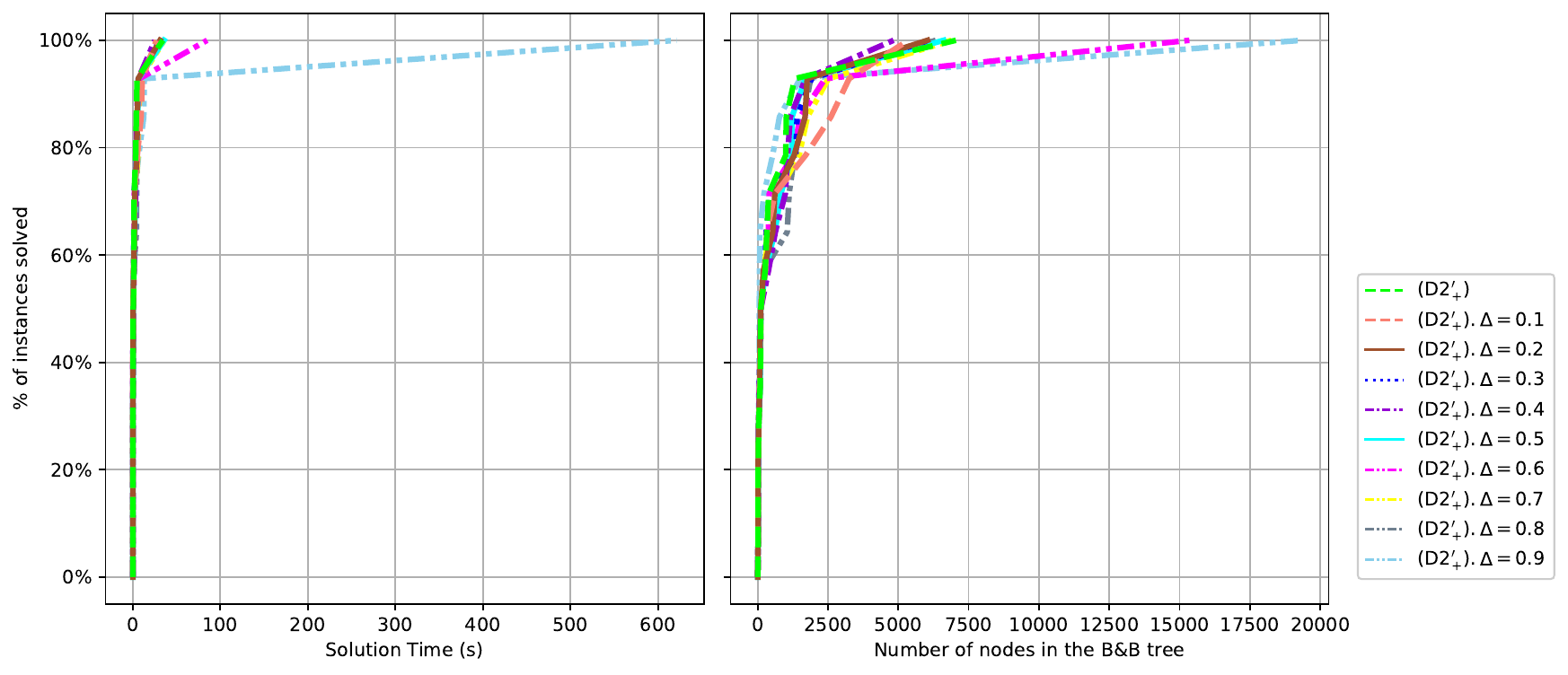}
    \caption{
    Time to solve the integer problem, and number of nodes in the B\&B tree, considering the effect of stabilization.
    }\label{fig:10}
\end{figure}

We also study the impact on stabilization in the performance of our branch-and-price. 
We conduct experiments employing different values of parameter $\Delta$ that represents the weight of the stabilized vector. Specifically we use $\Delta=[0.1, 0.2, ... , 0.8, 0.9]$, considering formulation $(\text{D2}_+^{\prime})$.
We observe in Figure \ref{fig:10} that using stabilization does not significantly impact the solution time positively. The value $\Delta=[0.9]$ in particular leads to an inefficient solving process. Further examination revealed that in some cases that takes longer, the main factor contributing to the algorithm's runtime appears to be the inefficiency in tightening the dual bound rather than issues with the primal bound.

In order to further accelerate the convergence in future research, we could study methods specifically aimed at improving the dual bound. Techniques such as advanced cut generation, tailored branching decisions focused, or specialized heuristics might be explored.

\begin{figure}[H] 
    \centering 
    \includegraphics[width=0.9\textwidth]{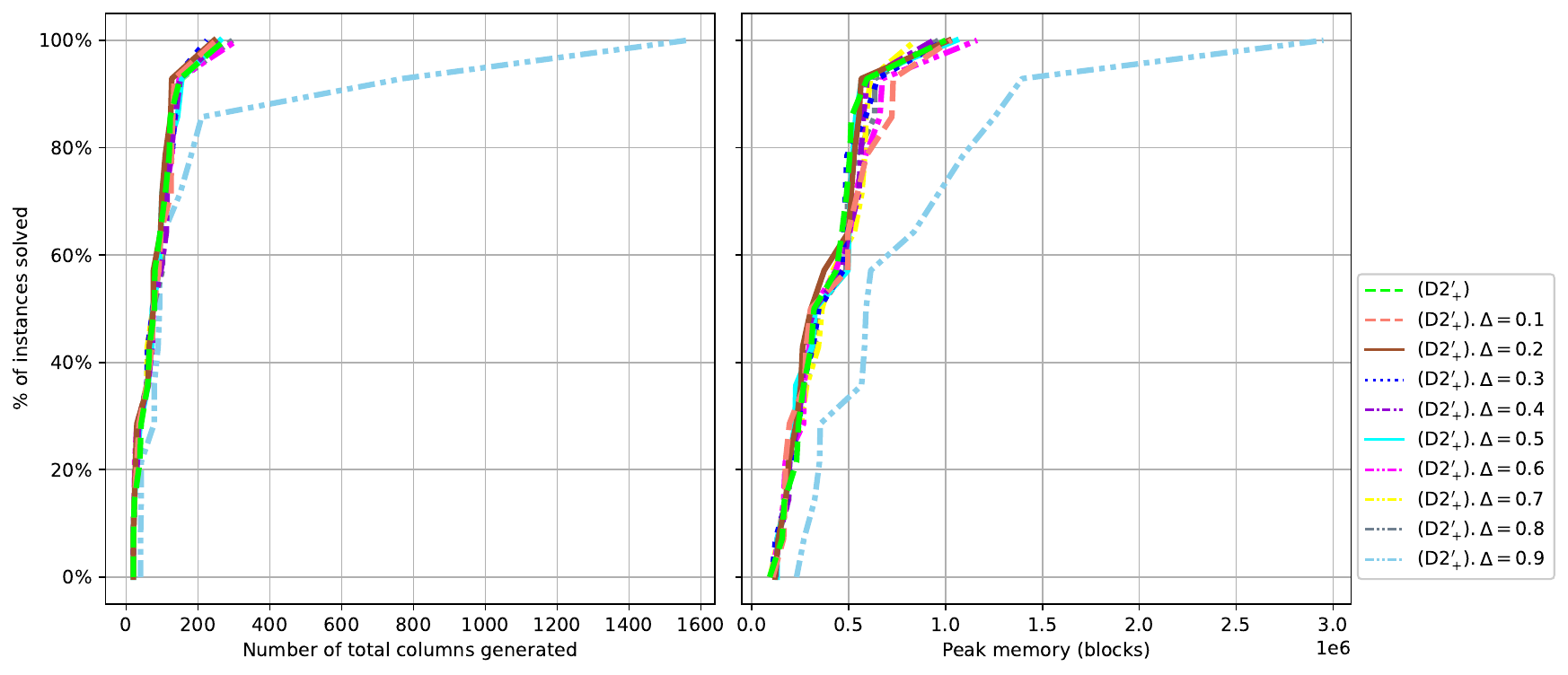}
    \caption{
    Number of total columns generated, and peak memory used to solve the integer problem considering the effect of stabilization.}\label{fig:11}
\end{figure}

On average, the number of columns generated is consistent across nearly all scenarios, with the exception of when $\Delta=0.9$. Additionally, similar to findings in previous experiments, there is a clear link between the number of columns generated and the peak memory usage (as indicated in Figure \ref{fig:11}).

\subsection{Comparison of {\normalfont$(\text{D2})$} with Benders and {\normalfont$(\text{D2}_+^{\prime})$} with branch-and-price} 

In noncompact Stackelberg games formulations, the number of variables can be intractable. To address this challenge, there are two common approaches. 
One method involves the application of the branch-and-price technique, while another possible approach is the utilization of the Benders technique.

To finish this section, we compare our $(\text{D2}_+^{\prime})$ approach with branch-and-price against a Benders implementation on $(\text{D2})$ where the variables $q$ are at the first \textcolor{black}{stage} of the Benders' structure, following the methodology of \citet{arriagada2021benders}. 

Benders technique needs to enumerate all the feasible knapsack strategies, and we associate each one of them with a different $x_i$ variable. Because of this, we work with smaller instances due to not being possible to solve larger instances within the time limit.  We consider $J = \{10,15\}$ and $K = \{2,4,6\}$.

\begin{figure}[H] 
    \centering \includegraphics[width=0.9\textwidth]{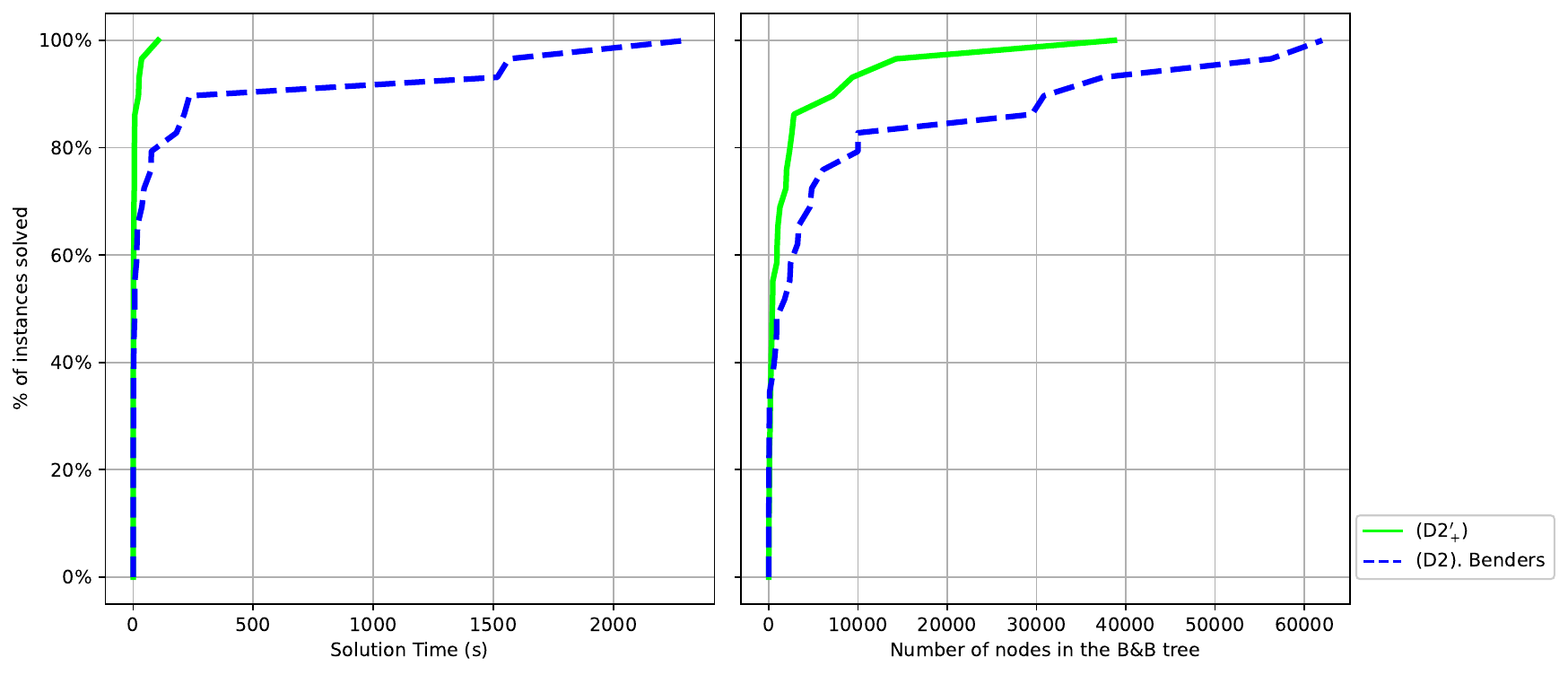}
    \caption{Time to solve the integer problem, and number of nodes in the B\&B tree, considering $(\text{D2})$ with Benders technique and $(\text{D2}_+^{\prime})$ with branch-and-price.}\label{fig:12}
\end{figure}

\begin{figure}[H] 
    \centering 
    \includegraphics[width=0.5\textwidth]{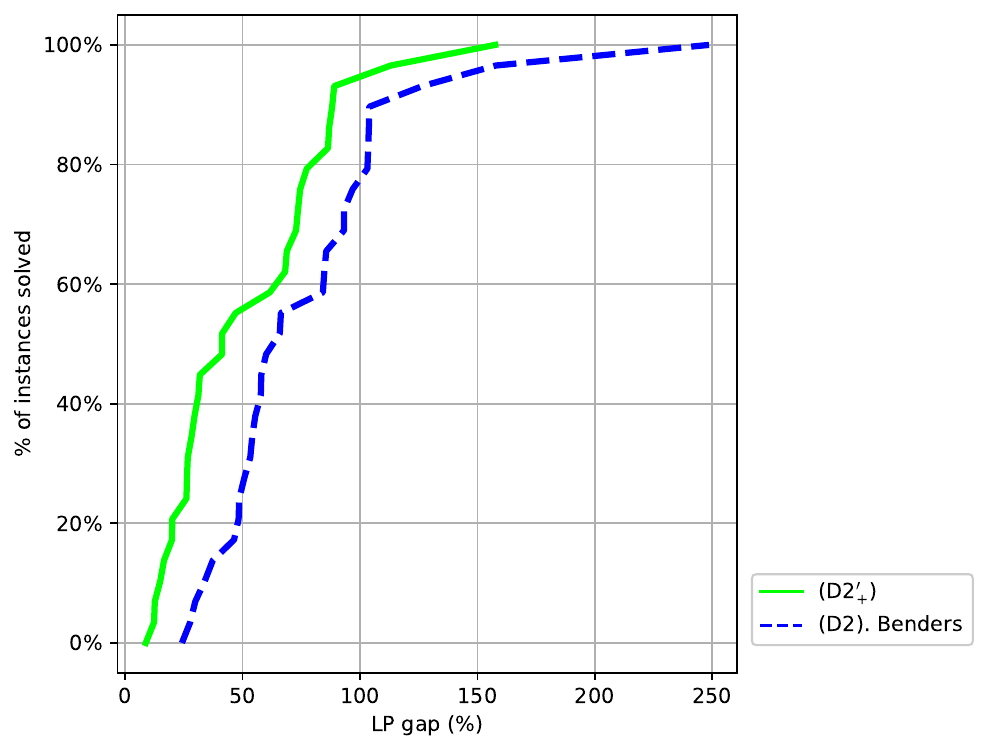}
    \caption{
    LP gap, considering $(\text{D2})$ with Benders technique and $(\text{D2}_+^{\prime})$ with branch-and-price.}\label{fig:13}
\end{figure}

In Figure \ref{fig:12}, we can observe that the Benders technique on $(\text{D2})$ is not as efficient as our $(\text{D2}_+^{\prime})$ with the branch-and-price approach. We can solve each of the instances of our $(\text{D2}_+^{\prime})$  formulation in less than \textcolor{black}{105s}, while with Benders,  we can solve only \textcolor{black}{80\%} of the instances in the same time frame. Also, as we escalate the size of the instance the problem gets more difficult to solve, as an example, it is possible to solve  \textcolor{black}{95\%} of the instances in less than 1500 seconds. Also note from Figure \ref{fig:13} that the LP Gap is approximately 25\% better when using the $(\text{D2}_+^{\prime})$ formulation.

\section{Conclusions}
\label{Conclusion}

We introduce novel valid inequalities in both Stackelberg games and Stackelberg security games, which define the new stronger formulations $(\text{D2}_+)$ and $(\text{D2}_+^{\prime})$, compared to $(\text{D2})$ formulation. Through the utilization of branch-and-price, we demonstrate the efficiency of our approach compared to the formulations in the literature, reducing the running time to 
\textcolor{black}{less than a fifth of the solution time} on a game consisting of protecting targets with different protection costs and a limited total budget.

We focus on improving (D2) formulation, as is the standard formulation for branch-and-price in Stackelberg Security Games. Its structure easily allows a generalized branch-and-price, unlike other formulations such as (Mip-k-G), which requires doing branch-and-cut-and-price, where the cuts are tailored for the specific game.

We evaluated our formulation using instances that vary in the number of targets, attackers, and budget. To the best of our knowledge, this is the first time that a study shows computational experiments using branch-and-price in the context of a Bayesian Stackelberg game, i.e., with multiple attackers.
 
From our experiments, we note a substantial improvement in the percentage of the LP gap for $(\text{D2}_+^{\prime})$ compared to $(\text{D2})$, leading to a higher-quality upper bound. This frequently translates into a faster solving process in the integer problem. Based on our experimental findings, we can conclude that as the number of targets and/or attackers increases, the $(\text{D2}_+^{\prime})$ formulation emerges as the fastest choice. Also, $(\text{D2}_+^{\prime})$ has less memory usage than $(\text{D2})$ and $(\text{D2}_+)$.

Also, our computational experiments explore the impact of diverse branch-and-price parameters on the efficiency of our method. 
Also, to enhance the efficiency of the branch-and-price algorithm, we found it is beneficial to begin with an initial subset of columns.  Based on empirical evidence, optimal mixed strategies typically utilize  $|J|$  or fewer strategies,  therefore we chose to work with $|J|$ initial columns, \textcolor{black}{unless stated otherwise}.
Within the context of the pricing problem, our experiments showed that it is more efficient to introduce just one strategy per iteration, since adding multiple strategies tends to slow down the computational process. Lastly, stabilization techniques did not significantly impact our solution time. Upon analyzing this, we found that the solution time seems to lie primarily in the dual bound rather than the primal.

 The utilization of branch-and-price will continue to be 
 a useful method for addressing Stackelberg Games, as they usually include an intractable number of strategies, which we can address using branch-and-price. Techniques to improve efficiency, such as improved column generation methods or improved branching strategies, remain an open area of research.

The insights obtained from this research can serve as a foundation for future work in Stackelberg Security Games and optimization techniques. With the newly introduced constraints, it is possible to address complex, real-world security scenarios more efficiently.
 
\newpage

\section{Acknowledgments}
This research was financially supported by the grant CONICYT-PFCHA/Doctorado Nacional/2019-21191904, the INRIA-Lille Programme Equipes Associées - BIO-SEL, the Open Seed Fund 2021 of Pontificia Universidad Católica de Chile and INRIA Chile, ANID PIA/PUENTE AFB230002, and FONDECYT-Chile, grant 1220047.  

\newpage

\bibliographystyle{apalike}
\bibliography{biblio}

\newpage
\begin{appendices}

\section{Infeasibility when branching on $q$}
\label{infeasibility}
As an example, consider a SSG with  $|J|=2$, $|K|=2$. We want to add columns iteratively through Branch-and-price. Let's suppose in our RMP we have $|I|=1$, where:
  \begin{align}
    P_1 &= \begin{bmatrix}
           1 \\
           0 
         \end{bmatrix}
  \end{align}
  
Lets suppose that we have as solution the following. If we consider all the possible defender strategies $P_i \, \forall i \in I$, our solution is:
  \begin{align}
&q_{2}^1=1 
  \end{align}

\begin{align}
\max \,& \sum_{k \in K} p^k f^k \\
\mbox{s.t } & f^k - (D^k(j|p) - D^k(j|u)) \sum_{i \in I}P^i_{j} x_i + Mq_{j}^{k} \le D^k(j|u) + M \; &\; & \forall j \in V, k \in K\\
&s^k - (A^k(j|p) - A^k(j|u)) \sum_{i \in I}P^i_{j} x_i + Mq_{j}^{k} \le A^k(j|u) + M  & \; & \forall j \in V, k \in K\\
&- s^k + (A^k(j|p) - A^k(j|u)) \sum_{i \in I}P^i_{j} x_i \le - A^k(j|u) \; &\; & \forall j \in V, k \in K\\
&\sum_{i \in I} x_i = 1 \; & \;&  \\
&\sum_{j \in V} q_j^k = 1& \;& \forall k \in K \\
&q_j^k \in \{0,1\}& & j \in V, k \in K\\
&s,f \in R^{\vert K \vert}&& \\
&x_i \ge 0&  & i \in I 
\end{align}

So we will enforce this solution with the only column $P$ we have. We will use $(\text{D2})$ formulation:
\begin{align}
\max \,& \sum_{k \in K} p^k f^k \\
\mbox{s.t } & f^k - (D^k(j|p) - D^k(j|u)) P^i_{j} + Mq_{j}^{k} \le D^k(j|u) + M \; &\; & \forall j \in V, k \in K\\
&s^k - (A^k(j|p) - A^k(j|u)) P^i_{j} + Mq_{j}^{k} \le A^k(j|u) + M  &(y_j^k) \; & \forall j \in V, k \in K\\
&- s^k + (A^k(j|p) - A^k(j|u)) P^i_{j} \le - A^k(j|u) \; &\; & \forall j \in V, k \in K\\
&\sum_{i \in I} x_i = 1 \; & \;&  \\
&\sum_{j \in V} q_j^k = 1& \;& \forall k \in K \\
&q_j^k \in \{0,1\}& & j \in V, k \in K\\
&s,f \in R^{\vert K \vert}&& \\
&x_i \ge 0&  & i \in I 
\end{align}

We extend our $\text{D2}$ model, considering the previously defined instance:
\begin{align}
\max \,\,& p^1f^1+p^2f^2 \\
\mbox{s.t } 
& f^1 \leq D^1(1|p) +M(1-q_{1}^{1})\; &\; & \\
& f^1 \leq D^1(2|u) + M(1-q_{2}^{1})\; &\; & \\
& s^1 \leq A^1(1|p) +M(1-q_{1}^{1})\; &\; & \\
& s^1 \leq A^1(2|u) + M(1-q_{2}^{1})\; &\; & \\
&s^1 \geq A^1(1|p) \; &\; & \\
& s^1 \geq  A^1(2|u) \; &\; & \\
&s^2 \geq A^2(1|p) \; &\; & \\
& s^2 \geq  A^2(2|u) \; &\; & \\
&x_1 = 1 \; & \;&  \\
&\sum_{j \in V} q_j^k = 1& \;& \forall k \in K \\
&q_j^k \in \{0,1\}& & j \in V, k \in K\\
&s,f \in R^{\vert K \vert}&& \\
&x_i \ge 0&  & i \in I 
\end{align}

\begin{align}
\max \,\,& p^1f^1+p^2f^2 \\
\mbox{s.t } 
& f^1 \leq D^1(1|p) +M\; &\; & \\
& f^1 \leq D^1(2|u) \; &\; & \\
& s^1 \leq A^1(1|p) +M\; &\; & \\
& s^1 \leq A^1(2|u) \; &\; & \\
&s^1 \geq A^1(1|p) \; &\; & \\
& s^1 \geq  A^1(2|u) \; &\; & \\
&x_1 = 1 \; & \;&  \\
&q_2^1 = 1& \;& \\
&q_j^k \in \{0,1\}& & j \in V, k \in K\\
&s,f \in R^{\vert K \vert}&& \\
&x_i \ge 0&  & i \in I
\end{align}

\begin{align}
\max \,\,& p^1f^1+p^2f^2 \\
\mbox{s.t } 
& f^1 \leq D^1(2|u) \; &\; & \\
& s^1 \leq A^1(2|u) \; &\; & \\
&s^1 \geq A^1(1|p) \; &\; & \\
& s^1 \geq  A^1(2|u) \; &\; & \\
&x_1 = 1 \; & \;&  \\
&q_2^1 = 1& \;& \\
&q_j^k \in \{0,1\}& & j \in V, k \in K\\
&s,f \in R^{\vert K \vert}&& \\
&x_i \ge 0&  & i \in I 
\end{align}

We finally have:
\begin{align}
\max \,\,& p^1f^1+p^2f^2 \\
\mbox{s.t } 
& f^1 \leq D^1(2|u) \; &\; & \\
& \bm{s^1 = A^1(2|u)} \; &\; & \\
&s^1 \geq A^1(1|p) \; &\; & \\
&x_1 = 1 \; & \;&  \\
&q_2^1 = 1& \;& \\
&q_j^k \in \{0,1\}& & j \in V, k \in K\\
&s,f \in R^{\vert K \vert}&& \\
&x_i \ge 0&  & i \in I 
\end{align}
Let us note that the bold expression states an equality, but this might be in conflict with the rest of the constraints regarding variable $s$.

This situation can be tackled by adding enough initial columns so that this situation does not arise. For example, this situation should not arise if we have one strategy per target, where only this said target is being protected. This is what we are currently doing, apart from also adding other columns obtained through an algorithm.

\newpage
\section{Proof for SSG strengthening constraints}
\label{proof_ssg}
\begin{proof}
If $q_j^k=1$:
\begin{align}
f^k &=  (D^k(j|p)-D^k(j|u)) \sum_{i \in I: j \in i} x_i + D^k(j|u) \\
&=  (D^k(j|p)-D^k(j|u))x_i +(D^k(j|p)-D^k(j|u)) \sum_{\substack{i^{\prime} \in I:\\ j \in i^{\prime}, i^{\prime} \neq i }} x_{i^{\prime}} + D^k(j|u) \\
&=  (D^k(j|p)-D^k(j|u))x_i +(D^k(j|p)-D^k(j|u)) \sum_{\substack{i^{\prime} \in I:\\ j \in i^{\prime}, i^{\prime} \neq i }} x_{i^{\prime}} + D^k(j|u) \nonumber \\ 
&+ (D^k(j|p)-D^k(j|u)) - (D^k(j|p)-D^k(j|u))\\
&=  -(D^k(j|p)-D^k(j|u))(1-x_i) +(D^k(j|p)-D^k(j|u)) \sum_{\substack{i^{\prime} \in I:\\ j \in i^{\prime}, i^{\prime} \neq i }} x_{i^{\prime}} + D^k(j|p) \\ 
&=  -(D^k(j|p)-D^k(j|u)) \sum_{\substack{i^{\prime} \in I:\\ i^{\prime} \neq i }} x_{i^{\prime}} +(D^k(j|p)-D^k(j|u)) \sum_{\substack{i^{\prime} \in I:\\ j \in i^{\prime}, i^{\prime} \neq i }} x_{i^{\prime}} + D^k(j|p) \\ 
&=  (D^k(j|p)-D^k(j|u)) (-\sum_{\substack{i^{\prime} \in I:\\ i^{\prime} \neq i }} x_{i^{\prime}} +\sum_{\substack{i^{\prime} \in I:\\ j \in i^{\prime}, i^{\prime} \neq i }} x_{i^{\prime}} )+ D^k(j|p) \\ 
&=  (D^k(j|p)-D^k(j|u)) (\sum_{\substack{i^{\prime} \in I:\\ i^{\prime} \neq i }} x_{i^{\prime}} (-1+P_j^{i^{\prime}})) + D^k(j|p) \\ 
f^k & \le   (\sum_{\substack{i^{\prime} \in I:\\ i^{\prime} \neq i }} x_{i^{\prime}} \max_{j \in J} (D^k(j|p)-D^k(j|u))(-1+P_j^{i^{\prime}})) + D^k(j|p) \\ 
& \le  \sum_{j \in J} D^k(j|p) q_j^k
\end{align}
\end{proof}
\end{appendices}

\newpage

\end{document}